\documentclass[]{spie}

\usepackage{amsmath,amsfonts,amssymb}
\usepackage{graphicx}
\usepackage[colorlinks=true, allcolors=blue]{hyperref}

\usepackage{multirow}
\usepackage{color}

\graphicspath{{./img/}}

\title{Optical modeling and polarization calibration for CMB measurements with
ACTPol and Advanced ACTPol}

\author[a]{Brian~Koopman}
\author[b]{Jason~Austermann}
\author[j]{Hsiao-Mei~Cho}
\author[g]{Kevin~P.~Coughlin}
\author[b]{Shannon~M.~Duff}
\author[a]{Patricio~A.~Gallardo}
\author[c,d]{Matthew~Hasselfield}
\author[a]{Shawn~W.~Henderson}
\author[f]{Shuay-Pwu~Patty~Ho}
\author[b]{Johannes~Hubmayr}
\author[j,k]{Kent~D.~Irwin}
\author[j]{Dale~Li}
\author[g]{Jeff~McMahon}
\author[e]{Federico~Nati}
\author[a]{Michael~D.~Niemack}
\author[l]{Laura~Newburgh}
\author[f]{Lyman~A.~Page}
\author[f]{Maria~Salatino}
\author[i]{Alessandro~Schillaci}
\author[e]{Benjamin~L.~Schmitt}
\author[f]{Sara~M.~Simon}
\author[a]{Eve~M.~Vavagiakis}
\author[e]{Jonathan~T.~Ward}
\author[h]{Edward~J.~Wollack}

\affil[a]{Department of Physics, Cornell University, Ithaca, NY 14853, USA}
\affil[b]{Quantum Devices Group, National Institute of Standards and
Technology, 325 Broadway M.S. 817.03, Boulder, CO 80305, USA}
\affil[c]{Department of Astronomy and Astrophysics, The Pennsylvania State
University, University Park, PA 16802, USA}
\affil[d]{Institute for Gravitation and the Cosmos, The Pennsylvania State
University, University Park, PA 16802, USA}
\affil[e]{Department of Physics and Astronomy, University of Pennsylvania, 209
South 33rd Street, Philadelphia, PA 19104, USA}
\affil[f]{Department of Astrophysical Sciences, Princeton University,
Princeton, NJ 08544, USA}
\affil[g]{Department of Physics, University of Michigan, Ann Arbor, MI 48109,
USA}
\affil[h]{NASA Goddard Space Flight Center, Greenbelt, MD 20771, USA}
\affil[i]{Instituto de Astrof\'isica and Centro de Astro-Ingenier\'ia, Facultad
de F\'isica, Pontificia Universidad Cat\'olica de Chile, Santiago, Chile}
\affil[j]{SLAC National Accelerator Laboratory, 2575 Sandy Hill Road, Menlo Park, CA 94025, USA}
\affil[k]{Department of Physics, Stanford University, Stanford, CA 94305-4085, USA}
\affil[l]{Dunlap Institute, University of Toronto, Toronto, Ontario M5S 3H4, Canada}

\authorinfo{Further author information: (Send correspondence to B.K.)\\B.K.:
E-mail: bjk98@cornell.edu, Telephone: 1 607 255 0833}

\usepackage{aas_macros}

\begin{document} 

\maketitle

\begin{abstract}
The Atacama Cosmology Telescope Polarimeter (ACTPol) is a polarization
sensitive upgrade to the Atacama Cosmology Telescope, located at an elevation
of 5190 m on Cerro Toco in Chile. ACTPol uses transition edge sensor bolometers
coupled to orthomode transducers to measure both the temperature and
polarization of the Cosmic Microwave Background (CMB).  Calibration of the
detector angles is a critical step in producing polarization maps of the CMB.
Polarization angle offsets in the detector calibration can cause leakage in
polarization from E to B modes and induce a spurious signal in the EB and TB
cross correlations, which eliminates our ability to measure potential
cosmological sources of EB and TB signals, such as cosmic birefringence.

We calibrate the ACTPol detector angles by ray tracing the designed detector
angle through the entire optical chain to determine the projection of each
detector angle on the sky. The distribution of calibrated detector polarization
angles are consistent with a global offset angle from zero when compared to the
EB-nulling offset angle, the angle required to null the EB cross-correlation
power spectrum. We present the optical modeling process.

The detector angles can be cross checked through observations of known
polarized sources, whether this be a galactic source or a laboratory reference
standard. To cross check the ACTPol detector angles, we use a thin film
polarization grid placed in front of the receiver of the telescope, between the
receiver and the secondary reflector. Making use of a rapidly rotating
half-wave plate (HWP) mount we spin the polarizing grid at a constant speed,
polarizing and rotating the incoming atmospheric signal. The resulting
sinusoidal signal is used to determine the detector angles.

The optical modeling calibration was shown to be consistent with a global
offset angle of zero when compared to EB nulling in the first ACTPol results
and will continue to be a part of our calibration implementation. The first
array of detectors for Advanced ACTPol, the next generation upgrade to ACTPol,
will be deployed in 2016. We plan to continue using both techniques and
compare them to astrophysical source measurements for the Advanced ACTPol
polarization calibration.
\end{abstract}


\section{INTRODUCTION}
\label{sec:intro}
The Atacama Cosmology Telescope (ACT) is an off-axis Gregorian telescope
constructed in 2007 \cite{2007ApOpt..46.3444F}. The Atacama
Cosmology Telescope Polarimeter (ACTPol) is a polarization sensitive
receiver upgrade to ACT. Starting in 2013, in a staged deployment, ACTPol
began observing the Cosmic Microwave Background (CMB).
In the most recent observation season ACTPol used ${\sim}3{,}000$ transition
edge sensor bolometers, in two frequency bands across three arrays
\cite{2016arXiv160506569T}.

\begin{figure} [ht]
\begin{center}
\begin{tabular}{c}
\includegraphics[width=0.5\linewidth]{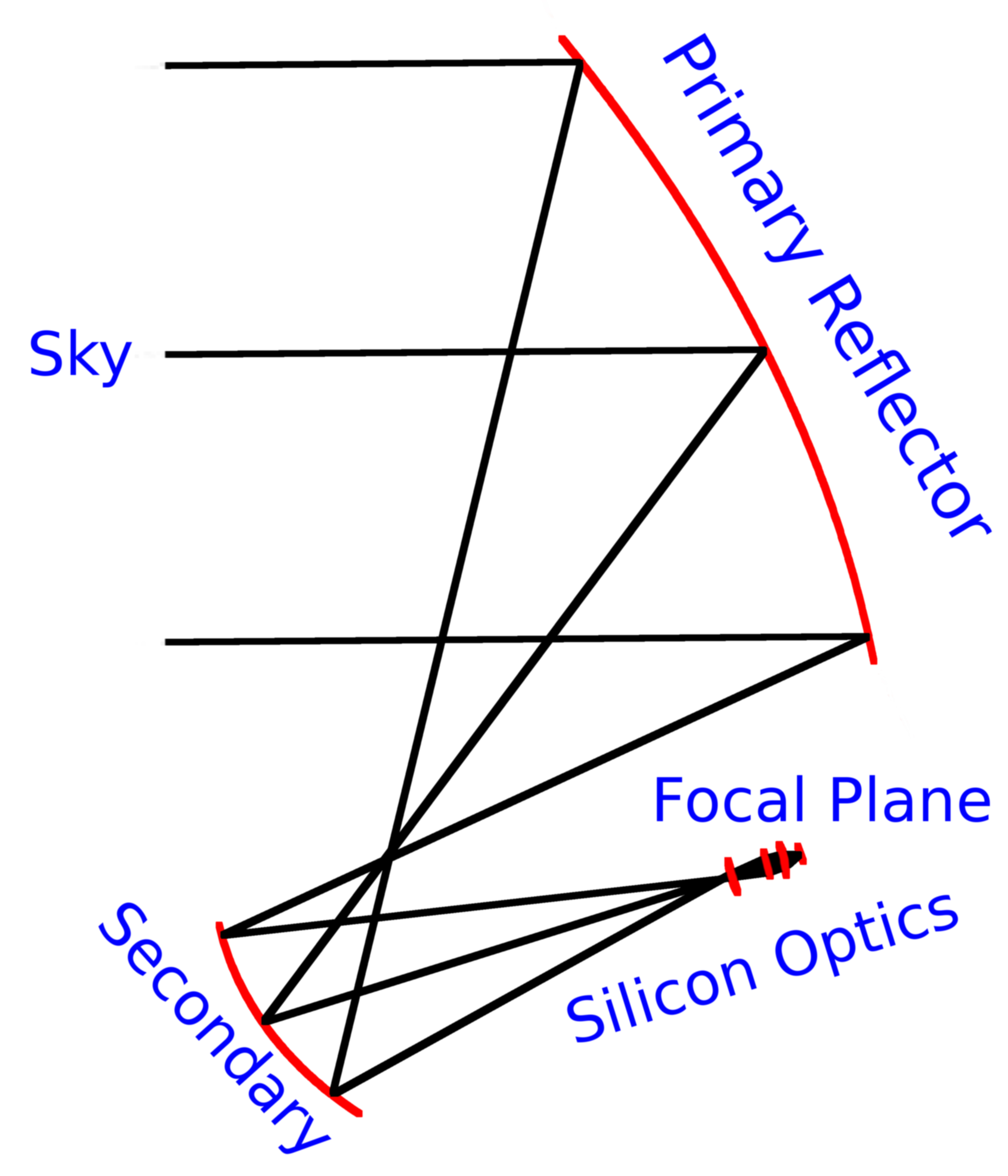}
\end{tabular}
\end{center}
\caption{\label{fig:ray_trace} 
Single field ray trace of the ACTPol optics.
Shown here is a simple ray trace from CODE V, with a single field point on the
sky propagated to the primary mirror, secondary mirror, and through the
refractive silicon optics in the ACTPol receiver for a single optics tube. The
focal plane for one array is the final element, where the ACTPol detectors
lie.}
\end{figure}

The primary elements of the ACTPol optical chain are a 6 m primary mirror, a 2
m secondary mirror and three optics tubes, each of which contains a set of
three silicon reimaging lenses. Other elements include the receiver window,
several band defining filters and a set of corrugated feedhorns per array.
Figure \ref{fig:ray_trace} shows the optical chain with a simple ray trace for
one optics tube.

The corrugated feedhorns each couple to an orthomode transducer
(OMT) which separates incoming light into orthogonal
polarizations. The angle of the two OMT fins are defined by lithography in
fabrication. The ACTPol detectors are fabricated on 3-inch wafers which
were etched into hexagonal and partial hexagonal shapes
(referred to as ``hexes'' and ``semi-hexes'') and tiled in an array of three
hexes and three semi-hexes each. The orthogonal
antenna probes on each wafer are oriented at 0/90 degrees and 45/135 degrees,
such that the full array of six wafers has sets of detectors ranging from 0 to
180 degrees at 15 degree intervals.

The initial set of calibration angles are determined by the
photolithography of the detector wafers and their placement into the array. We
then add an array-specific angle when the array is installed into the receiver
cryostat. This angle is constrained mechanically, hereafter referred to as the
``installation angle.'' Finally, the optics chain itself (the reflectors and
lenses) causes a position dependent polarization rotation, which we solve for using the
optical design software CODE V\footnote[1]{Synopsys Optical Solutions Group --
https://optics.synopsys.com/codev/}. All three of these angles are combined to
produce the final polarization angle calibration for ACTPol.

\section{Polarization Calibration and Cosmic Polarization Rotation}
Cosmic polarization rotation (CPR), also referred to as
``cosmological birefringence,'' is the rotation of linearly
polarized light as it traverses empty space at cosmological scales
\cite{PhysRevLett.83.1506, 1998PhRvL..81.3067C}. The search for a non-zero CPR
is a test for CPT-violating physics and
has been probed across the electromagnetic spectrum with observations of
polarized radio and UV emission from galaxies, polarized gamma ray bursts and
the CMB polarization signal \cite{1998PhRvL..81.3067C, 1997astro.ph..4285L,
PhysRevLett.79.1801, 2014MNRAS.444.2776G, 2011ApJS..192...18K}.

The best constraints on CPR today come from observations of the polarization of
the cosmic microwave background (CMB) \cite{2016MNRAS.455.1981K}. The CMB can
be used to put constraints on CPR through the cross-correlation power spectra.
CMB temperature and polarization maps can be expanded in the spherical
harmonics $Y_{\ell m}^X$. The covariances of the corresponding expansion
coefficients $a^X_{\ell m}$ form the angular power spectra,
\begin{equation}
C_{\ell}^{XX} \delta_{\ell \ell'} \delta_{mm'} = \langle a_{(\ell m)}^{X*} a_{(\ell'm')}^{X} \rangle,
\end{equation}
where $XX$ can be any combination of $T$, $E$, $B$ (e.g. $TT$, $EE$, $BB$, $TE$,
$TB$, $EB$) \cite{PhysRevLett.78.2058, 1997PhRvD..55.1830Z}. The distribution of inhomeogeneities in the CMB are
required to be invariant under parity and since $Y^T_{\ell m}$ and $Y^E_{\ell m}$
have parity $(-1)^{\ell}$ and $Y^B_{\ell m}$ has parity $(-1)^{\ell+1}$ we expect
$C_{\ell}^{TB} = 0$ and $C_{\ell}^{EB} = 0$ \cite{PhysRevLett.78.2058}. A
non-zero CPR angle would cause a mixing of E-modes into
B-modes, resulting in non-zero EB and TB cross-correlation power spectra
($C^{'TB}_{\ell}$, $C^{'EB}_{\ell} \neq 0$) \cite{PhysRevLett.83.1506}.
The amount of mixing caused by a CPR rotation,
$\alpha$, is given as \cite{PhysRevLett.83.1506, 2013ApJS..208...19H, 2013ApJ...762L..23K},
\begin{align}
C^{'TE}_{\ell} &= \cos \left( 2 \alpha \right) C_{\ell}^{TE}, \label{eqn:TE}\\
C^{'EE}_{\ell} &= \sin^2 \left( 2 \alpha \right) C_{\ell}^{BB} + \cos^2 \left( 2 \alpha \right) C_{\ell}^{EE},\\
C^{'EB}_{\ell} &= \frac{1}{2} \sin \left( 4 \alpha \right) \left( C_{\ell}^{BB} - C_{\ell}^{EE} \right), \label{eqn:EB}\\
C^{'TB}_{\ell} &= - \sin \left( 2 \alpha \right) C_{\ell}^{TE}, \label{eqn:TB}\\
C^{'BB}_{\ell} &= \cos^2 \left( 2 \alpha \right) C_{\ell}^{BB} + \sin^2 \left( 2 \alpha \right) C_{\ell}^{EE}. \label{eqn:BB}
\end{align}
Here the primes on the left side indicate the observed, rotated, power spectra.
The non-primed terms give the original, unrotated, quantities. Looking at
Equations \eqref{eqn:EB} and \eqref{eqn:TB}, we see that a non-zero $EB$ and
$TB$ power spectrum can result from a mixing of $E$ into $B$ modes.

Any miscalibration in the detector angles of an experiment is degenerate with
a non-zero CPR. As a result, current constraints on $\alpha$ are limited by the
systematic error associated with calibrating the polarization angle of the
detectors. Current polarization techniques limit the angular resolution to
roughly $\pm 0.5^{\circ}$ \cite{2016MNRAS.455.1981K}. In addition, using the assumption that $C^{'TB}_l$
and $C^{'EB}_l = 0$ to determine the calibration angle eliminates the ability
to measure a DC CPR.

\section{OPTICAL MODELING CALIBRATION}
\label{sec:optical_modeling}
The ACTPol telescope and receiver design was modeled with the optical design
software CODE V. Using CODE V we model the
ACTPol telescope, perform a ray trace to construct a function which transforms
focal plane coordinates to coordinates on the sky, and finally use a
polarization sensitive ray trace to calculate polarization rotations caused by
the optics. Together, they provide a
final polarization angle calibration for each detector on the focal plane,
which forms a critical input to the map making process.

\subsection{Ray Tracing}
\label{sec:ray_tracing}
As described in Section \ref{sec:intro}, the starting point for our calibration
is the lithographically defined angles for each detector. Tiled into an array
these populate every 15 degrees from 0 to 180 degrees. Table \ref{table:angles}
shows the distribution of angles for each array. The tiled detectors all lie
behind a nanofabricated set of corrugated feedhorns whose positions are
known precisely, which establishes a set of focal plane coordinates. These
focal plane coordinates, coupled with the detector angles, form our initial
parameters. All rotations are applied to these angles to produce our final
angle calibration.

\begin{table}[h!]
\caption{Distribution of detector angles in focal plane coordinates for all three ACTPol arrays.}
\label{table:angles} 
\begin{center}
\begin{tabular}{|c|c|c|c|}
\hline
Angle & PA1 & PA2 & PA3\\
\hline
$0^{\circ}$ & 87 & 80 & 86 \\
\hline
$15^{\circ}$ & 87 & 87 & 76 \\
\hline
$30^{\circ}$ & 80 & 87 & 86 \\
\hline
$45^{\circ}$ & 87 & 79 & 84 \\
\hline
$60^{\circ}$ & 87 & 87 & 78 \\
\hline
$75^{\circ}$ & 79 & 87 & 84 \\
\hline
$90^{\circ}$ & 87 & 80 & 86 \\
\hline
$105^{\circ}$ & 87 & 87 & 76 \\
\hline
$120^{\circ}$ & 80 & 87 & 86 \\
\hline
$135^{\circ}$ & 87 & 79 & 84 \\
\hline
$150^{\circ}$ & 87 & 87 & 78 \\
\hline
$165^{\circ}$ & 79 & 87 & 84 \\
\hline
\end{tabular}
\end{center}
\end{table}

The first step in our calibration is to determine the installation angle,
which is a global rotation of the detectors as an array is installed into the
ACTPol receiver. This installation angle is different for each array and
arises due to mechanical constraints within the receiver. Observations of
Saturn and Uranus are fit to a 2D Gaussian per detector, which gives
us pointing information for each feed and allows us to determine the
installation angle. Knowing where each detector is in sky coordinates allows
one to construct a scaling to take the focal plane coordinates to sky
coordinates. A global rotation can then be applied to match the observed sky
coordinates.  This global rotation does not take into account smaller
position and rotation angle distortions caused by the optical chain. We
can recover these optical distortions with CODE V.

CODE V is an optical ray trace code that we use to trace the sky through
the optics to the focal plane. CODE V allows a user to set up to 25 input
fields at a time. The code traces each of the 25 input sky fields through
the reflectors and lenses back to the feed horns. The final ray trace
provides a mapping between the coordinates on the sky and position on the focal
plane. The point spread function (PSF) for each field is then computed,
returning the PSF centroid relative to the chief ray coordinates returned by
the real ray trace. The locations of each PSF centroid are combined with the
real ray trace results to form a final focal plane coordinate per input field.
The 25 fields are then fit to a 2D quadratic in $x$ and $y$, shown as
$f(x, y)$ in Equation (\ref{eqn: quad_fit}), where $x$
and $y$ are in focal plane coordinates. This produces two functions to
transform focal plane coordinates to sky coordinates, one for $x_{sky}$ and one
for $y_{sky}$.
\begin{equation}
f(x, y) = A + Bx + Cy + D x^2 + E x y + F y^2
\label{eqn: quad_fit}
\end{equation}
These fits are then used to propagate each feedhorn
location to the sky. This incorporates any optical distortions on the positions
caused by the ACTPol optics to 2nd order and is a clear improvement over
a simple scaling. Given the focal plane coordinates for each feedhorn in an
array we then propagate those coordinates to the sky and perform a least
squares minimization, allowing a global translation and rotation, to
match the modeled sky coordinates to those determined for each feedhorn by
planet observations.

\begin{figure} [ht]
\begin{center}
\begin{tabular}{cc}
\includegraphics[width=0.5\linewidth]{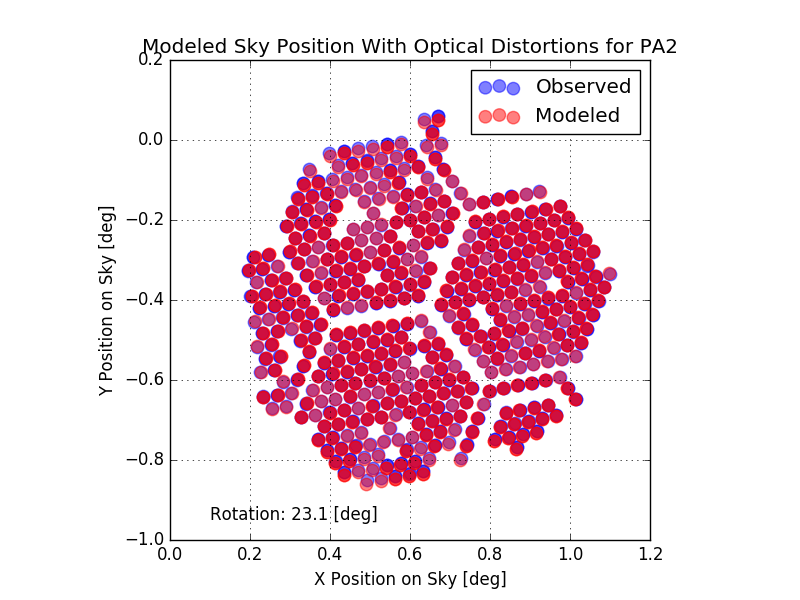} & \includegraphics[width=0.5\linewidth]{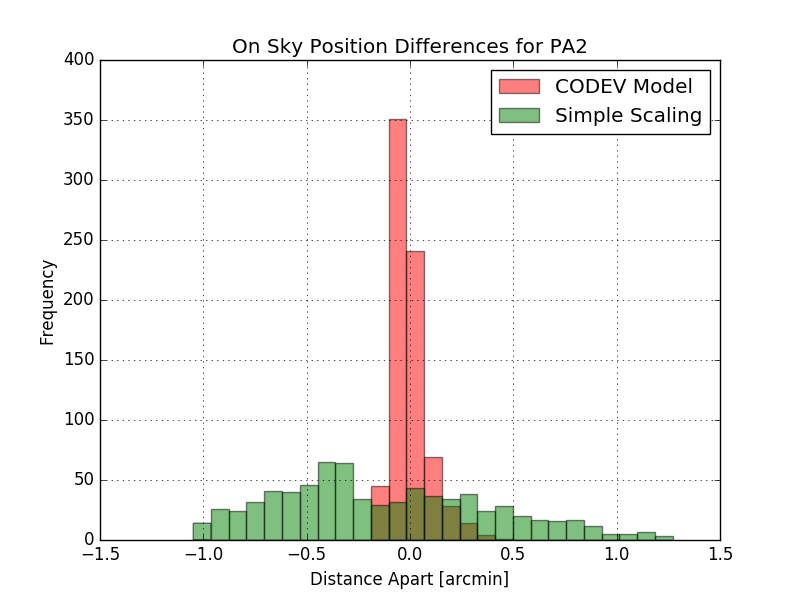}\\
\end{tabular}
\end{center}
\caption{\label{fig:obs_sky_pos} 
(Left): Feedhorn locations in sky coordinates as determined by planet
observations (Observed in blue) and by ray tracing in CODE V (Modeled in red).
The CODE V modeling locations are computed using the fit to the ray trace and
point spread functions given by Equation (\ref{eqn: quad_fit}). (Right):
Histogram of differences between observed and modeled detector positions on the
sky, calculated for a simple scaling to transform focal plane coordinates
to sky coordinates (green) and for the CODE V modeled ray trace (red).}
\end{figure}

Figure \ref{fig:obs_sky_pos} (left) shows a plot of the PA2 focal plane
coordinates propagated to sky coordinates using the fit to the CODE V model
along with the determined detector positions from planet observations. The
least squares minimization has been performed and for PA2 returns a required
rotation of 23.1 degrees from our initial focal plane coordinates for that
array. This forms our installation rotation and is applied globally to all
detector angles in the array. Figure \ref{fig:obs_sky_pos} (right) shows
a histogram of the differences in position per feed horn after least squares
minimization for PA2 using a simple scaling versus using the CODE V model,
showing we can model the position angle of each detector to better than
$20''$.

\subsection{Polarization Rotation}
The next step in the calibration is to calculate polarization rotation caused
by the ACTPol optics. CODE V is capable of performing a polarization sensitive
ray trace that can compute the transmittance at each surface from the
Fresnel equations. We define the input polarization field to be
identical for each sky trace and then use the poldsp
MACRO\footnote[3]{poldsp is a user supplied MACRO written by members of the
Physics Department at the University of Alabama, Huntsville which supplements
the polarization output from CODE V.} to calculate the polarization state for
nineteen rays across the entrance pupil diameter. The average polarization
across the entrance pupil diameter is then computed per field.

\begin{figure} [ht]
\begin{center}
\begin{tabular}{c}
\includegraphics[width=0.5\linewidth]{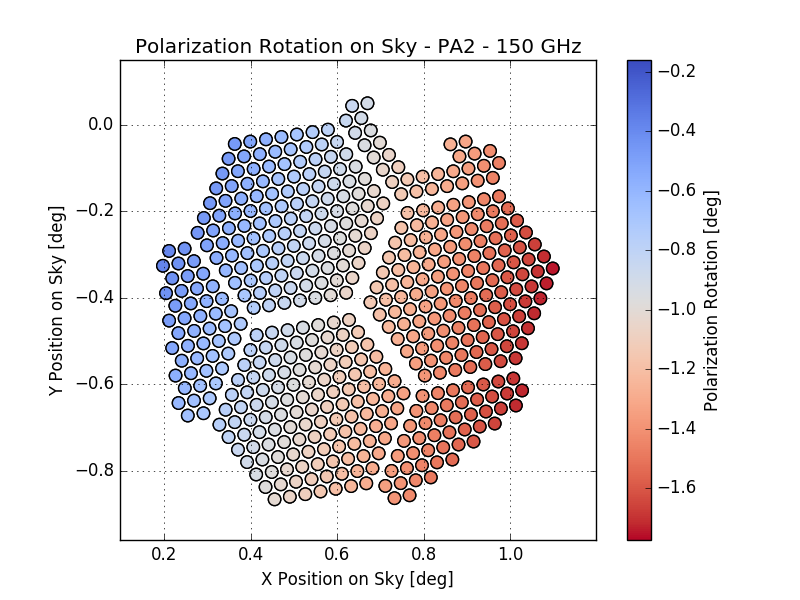}\includegraphics[width=0.5\linewidth]{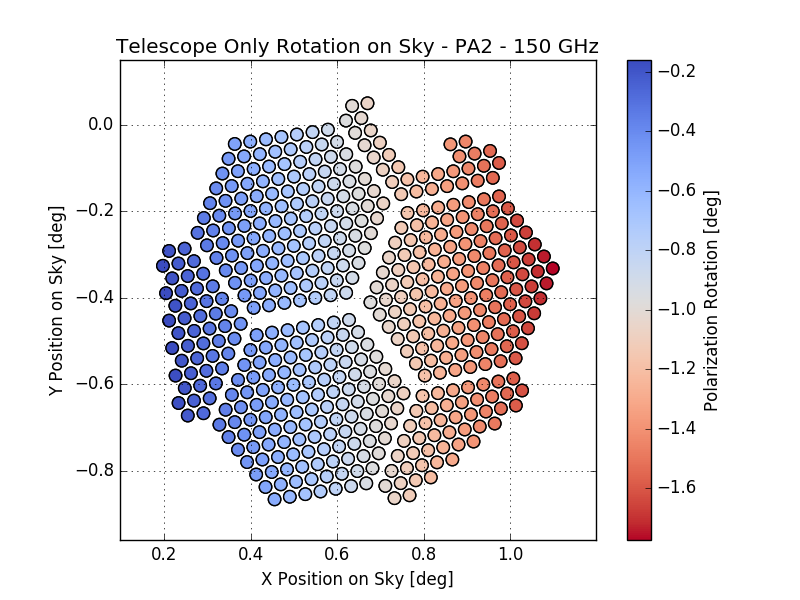}
\end{tabular}
\end{center}
\caption{\label{fig:pol_telescope_focal} 
(Left): Polarization rotation across the PA2 focal plane for the entire optical
chain at 146 GHz as determined by CODE V, plotted in sky coordinates. (Right):
Polarization rotation from the ACTPol reflectors only. The final surface in this
calculation is the focus. Note that the reimaging optics introduce a rotation.}
\end{figure}

This set of polarization rotations is used in another 2D quadratic
fit using Equation (\ref{eqn: quad_fit}), similar to what was done to transform
the focal plane coordinates to sky coordinates. The fit results are then
used to calculate the rotation associated with each feedhorn position on the
focal plane. This is performed for the entire optical chain, producing the
polarization rotations for each feedhorn, which are then combined with
the initial angle and installation angle to produce a final calibration angle
per detector. The left side of Figure \ref{fig:pol_telescope_focal} shows the
resulting polarization rotation across the PA2 focal plane plotted in sky
coordinates. To show the distribution of rotations due to just the telescope
mirrors the right side of Figure \ref{fig:pol_telescope_focal} shows the
polarization rotation at the telescope focus for PA2.

Accounting for the anti-reflection coatings on the ACTPol optics is important in
properly modeling the polarization rotation. The lenses are off-axis and
thus influence the propagation of the polarization vectors through the
camera according to their Fresnel coefficient. ACTPol uses multilayer
metamaterial anti-reflection (AR) coatings on each lens in the optical chain.
\cite{2013ApOpt..52.8747D} These multilayer coatings are
entered in CODE V using a multilayer coating definition file (MUL),
defining their thickness and index of refraction. For the ACTPol
coatings, we used physically measured thicknesses and HFSS simulated
effective indices for the coating parameters. These parameters are shown in
Table \ref{table:mul_params}.

\begin{table}[ht]
\caption{Coating parameters used to model the 2-layer and 3-layer anti-reflection
coatings applied to the ACTPol lenses. The index of refraction for the silicon
substrate is modeled to be 3.384 for all lenses \cite{2013ApOpt..52.8747D}.}
\label{table:mul_params} 
\begin{center}
\begin{tabular}{c|c|c|}
\cline{2-3}
& \textbf{Thickness [$\mu$m]} & \textbf{Index}\\
\hline
\multicolumn{1}{|c|}{\multirow{2}{*}{2-Layer}} & 365 & 1.38\\
\multicolumn{1}{|c|}{} & 200 & 2.50\\
\hline
\multicolumn{1}{|c|}{\multirow{3}{*}{3-Layer}} & 470 & 1.28\\
\multicolumn{1}{|c|}{} & 315 & 1.95\\
\multicolumn{1}{|c|}{} & 245 & 2.84\\
\hline
\end{tabular}
\end{center}
\end{table}

\begin{figure} [ht]
\begin{center}
\begin{tabular}{c}
\includegraphics[width=0.75\linewidth]{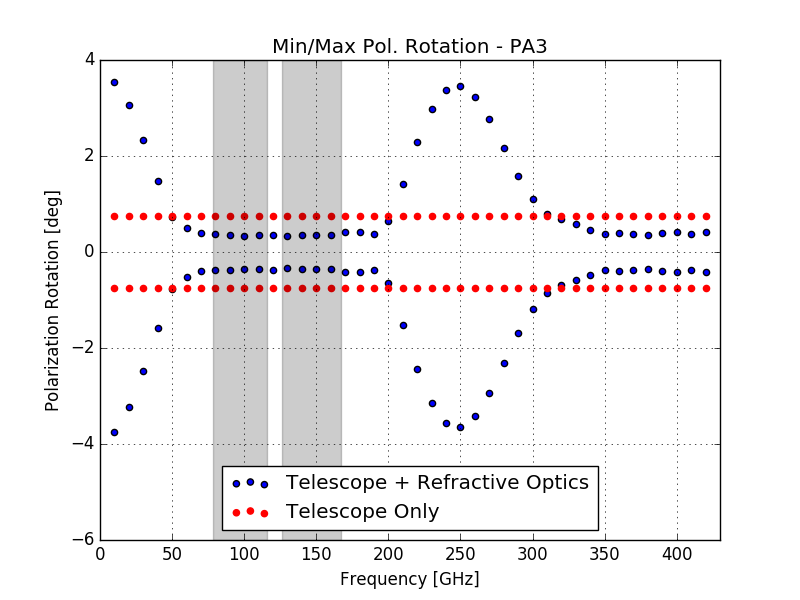}
\end{tabular}
\end{center}
\caption{\label{fig:min_max_pa3} 
Plot of the minimum and maximum polarization rotations caused by the ACTPol
optical chain as a function of frequency of the incoming light. Plotted in red
are the minimum and maximum polarization rotation contributions from the
telescope mirrors only, terminating at the receiver window. Plotted in blue are
the minimum and maximum polarization rotations for the entire optical chain,
including the anti-reflection coated silicon refractive optics. Highlighted in
gray are the upper and lower science bands for CMB observations with PA3.
Polarization rotation is similarly uniform across the science bands for PA1 and
PA2.}
\end{figure}

Calculating the polarization rotation for many different wavelengths across the
ACTPol science band shows the wavelength dependence of the rotations caused by
the optics. Figure \ref{fig:min_max_pa3} shows this for PA3, the multi-chroic
array in ACTPol. The polarization rotation is uniform across frequencies within
the highlighted science bands, where the AR coating parameters are tuned for
maximum transmission. Figures \ref{fig:pol_telescope_focal} and
\ref{fig:min_max_pa3} show that the refractive optics reduce the total range of
polarization rotations in band across the focal plane.

\subsection{Optical Modeling Results}
\begin{figure} [ht]
\begin{center}
\begin{tabular}{c}
\includegraphics[width=1.0\linewidth]{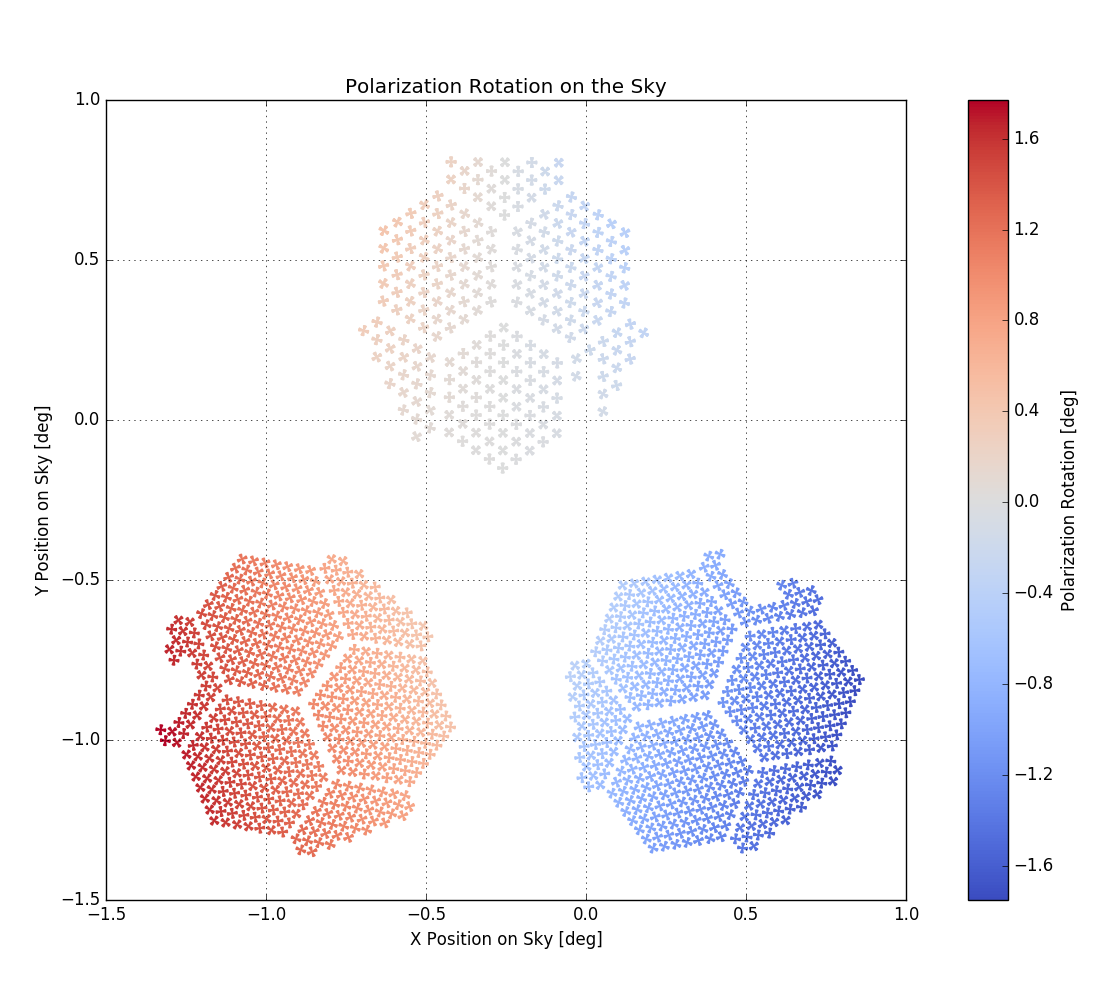}
\end{tabular}
\end{center}
\caption{\label{fig:full_calibration} 
Plot of all three ACTPol arrays in sky coordinates. The angles of the plotted
points correspond to the physical angle of the detectors in the telescope. The
color scale corresponds to the additional polarization rotation caused by the
ACTPol optics that is required to complete the angle calibration.  This
additional -1.7 to 1.7 degree rotation has been an important input to the
detector angle calibration in published ACTPol results.}
\end{figure}

Working from the initial, lithographically defined, detector positions and
polarization angles we apply the series of calibration rotations detailed above
to form a final polarization angle calibration per detector for making maps of
the CMB polarization. The final ACTPol detector angle calibrations
are shown in Figure \ref{fig:full_calibration}, with colors and
marker shape indicating with and without the CODE V correction for the
polarization angle.

This CODE V angle, while small, has been critical in calibrating
the detectors for ACTPol. Results published from the first season of
ACTPol data used polarization angles derived from the prescription outlined in
this proceedings, and checked the results by measuring the EB cross-spectrum.
The EB nulling angle was found to be consistent with zero, $\delta \gamma_p =
-0.2^\circ \pm 0.5^\circ$, implying that the polarization angle
calibration by optical modeling works at the $0.5^\circ$
level or better.\cite{2014JCAP...10..007N}

\section{POLARIZATION GRID CALIBRATION}
We built a thin film polarization grid to directly measure the
polarization rotation caused by the optics. The polarization grid was placed at
the entrance to the receiver and used to polarize the input to the receiver.
This procedure is used to extract a per detector polarization angle and is
compared to the optics only polarization rotation as determined in CODE V.

\subsection{Polarization Grid Design}
\label{sec:wire_grid}
A large, rapidly modulating, polarized signal on the detectors would drive
the detectors into a nonlinear regime. With this in mind we explored different
polarization grid geometries using CST
Microwave Studio\footnote[4]{https://www.cst.com/products/cstmws}, varying the
wire pitch and width for a given material thickness to produce a mostly
reflective grid.  This would produce a small polarized signal on the detectors
so that they remain stable during measurements.

The grid is a thin metal film on a PolyEthylene Terephthalate (PET) substrate,
$12\,\mu\mathrm{m}$ thick (48 gauge), sourced from Dunmore
Corporation\footnote[2]{http://www.dunmore.com}. The PET is fully
metalized on one side with $30\,\mathrm{nm}$ of aluminum. The desired geometry
is then made through laser ablation of the aluminum layer by PhotoMachining
Inc.\footnote[1]{http://photomachining.com} Based on simulations in CST
Microwave Studio, we chose a geometry of straight, parallel wires,
$950\,\mu\mathrm{m}$ wide, on a $1000\,\mu\mathrm{m}$ pitch.

This produces a grid with 97\% reflection in one
polarization and 99.8\% in the other polarization at $150\,\mathrm{GHz}$, as
determined by CST Microwave Studio. We measured the reflectance of the grids to
be $95 \pm 5$\% reflective in one polarization, $90 \pm 7$\% in the other
polarization at 150 GHz using a custom reflectometer setup at the
University of Michigan.

\subsection{Polarization Grid Assembly and On Telescope Performance}
\begin{figure} [ht]
\begin{center}
\begin{tabular}{cc}
\includegraphics[width=0.5\linewidth]{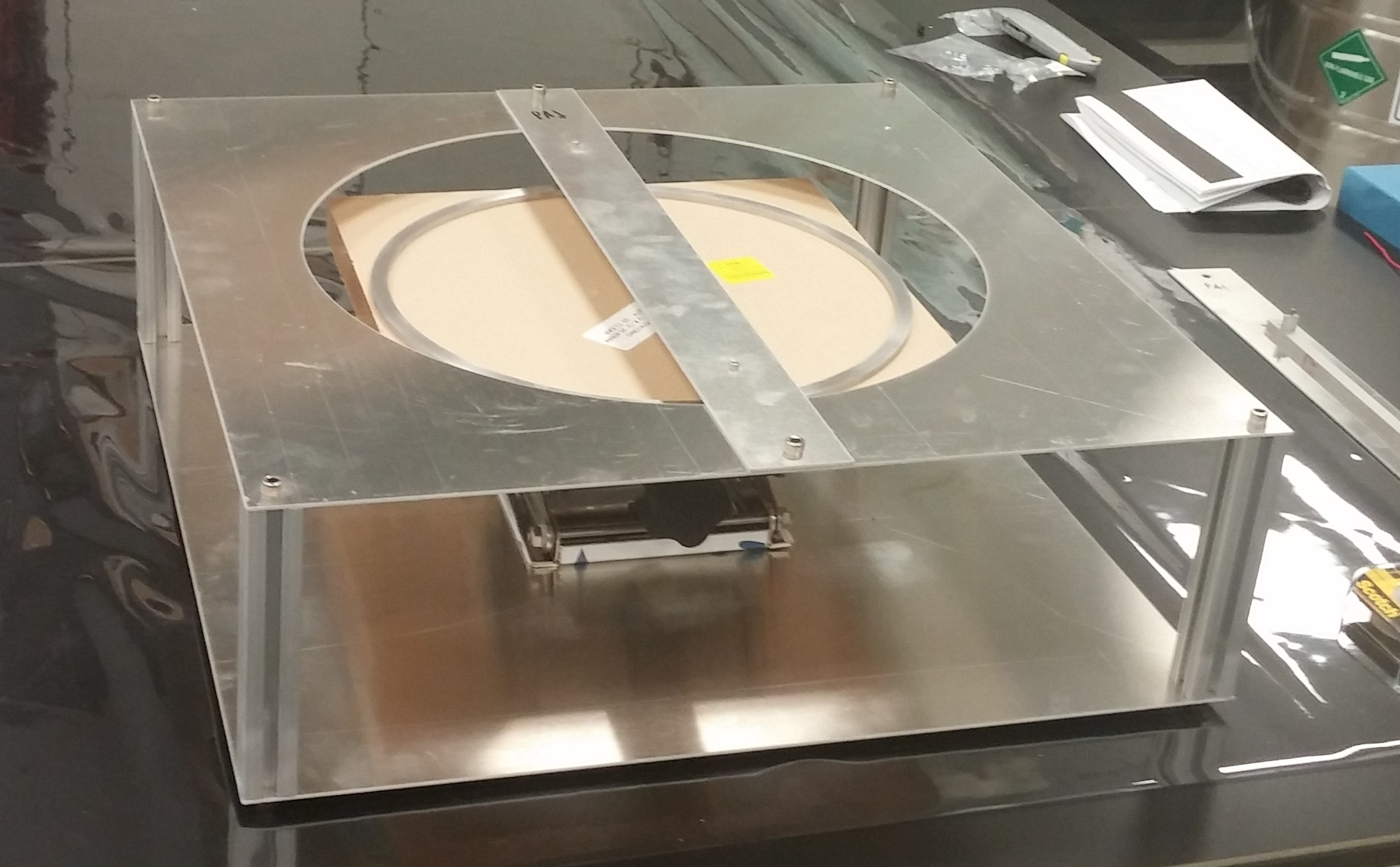} & \includegraphics[width=0.4\linewidth]{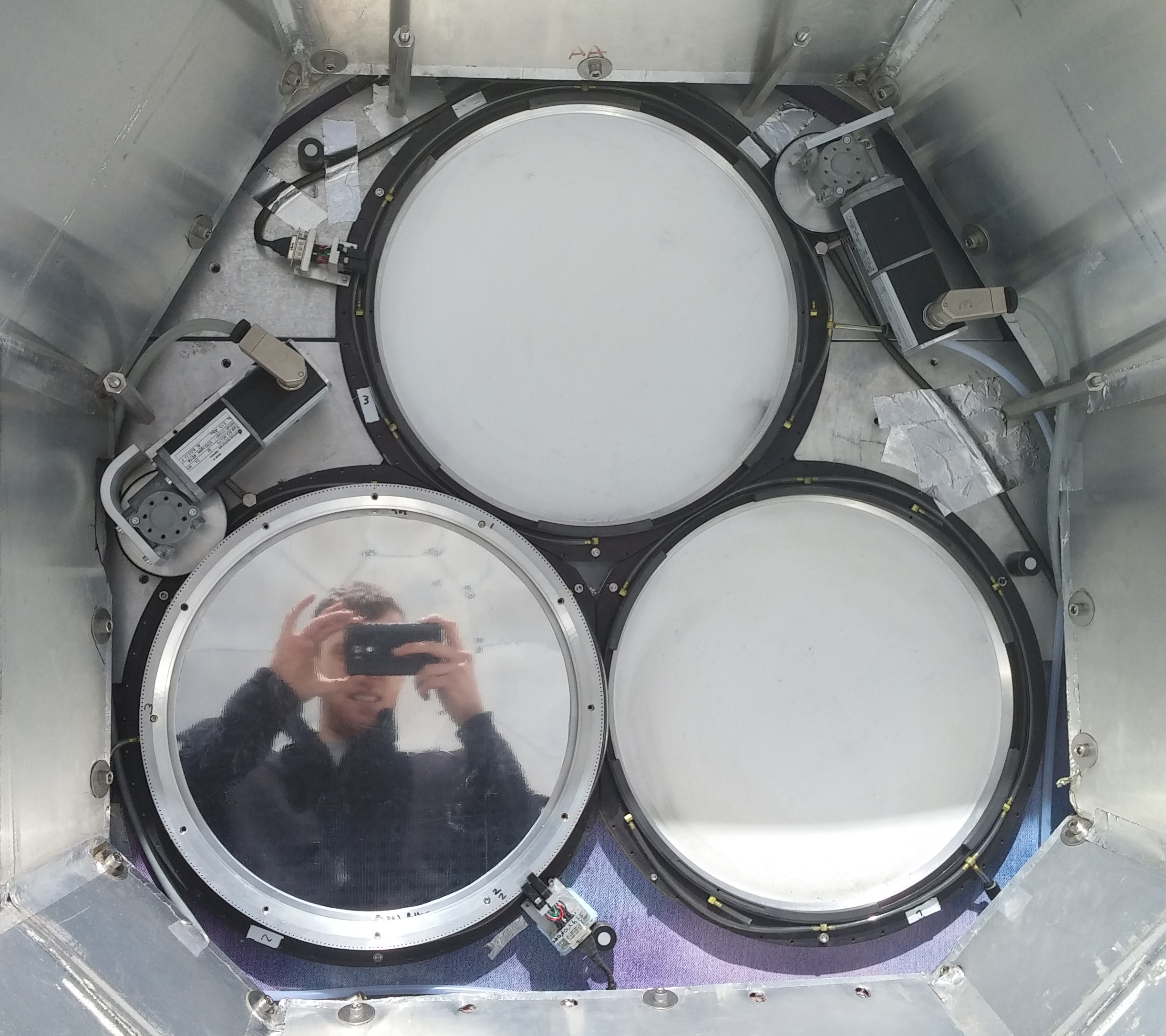}\\
\end{tabular}
\end{center}
\caption{\label{fig:grid_photo} 
(Left): Polarization grid mounting and alignment stand. The
polarization grid is taped taut across the round opening
and aligned using the pictured cross bar using alignment
marks on the grid to alignment holes on the aluminum mounting ring. The
aluminum ring is epoxied and raised into the grid, making it taut and smooth.
The epoxy is allowed to set and then excess material is removed. (Right): Photo
of the front of the ACTPol receiver. The arrays are numbered 1-3 starting with
the bottom right and moving clockwise. As pictured here PA1 and PA3 have
nothing installed in front of them, PA2 has the thin film polarizing grid
installed. The motor seen up and to the left of the polarization grid on PA2 rotates
the grid which is floating on an air bearing. The encoder, in the bottom right,
and the associated readout electronics (not pictured) record the angular
position of the grid during rotation.}
\end{figure}

The polarization grid was attached to a simple aluminum
ring, the same thickness as the planned Advanced ACTPol half-wave plates,
to fit into the half-wave plate mounting hardware. We built a simple
mounting stand to attach the grid to the aluminum ring. Figure
\ref{fig:grid_photo} (left) shows the polarization grid stand; the grid
was securely attached to an aluminum panel with a circular cutout. The aluminum
ring was then epoxied and raised into the PET side of the
polarization grid, providing tension while the epoxy
set. Two polarization grids were successfully mounted
using the test stand, one for each of the two 150 GHz optics tubes on ACTPol.

The polarization grids were individually installed
into the Advanced ACTPol half-wave plate mounts on the front of the telescope
receiver. The half-wave plate mounting hardware contains an
air bearing, allowing the mounted polarization grids to
spin friction free. Coupled with an external motor the
polarization grids were spun at 0.5, 1 and 2 Hz during
measurements. Several sets of measurements were performed at these three
constant rotational rates, rotating both clockwise and counter clockwise.
During measurements, the telescope remained stationary.

The right panel of Figure \ref{fig:grid_photo} shows one of the polarization
grids installed on PA2, the second ACTPol array. With the grid rotating at a
constant rate, the detectors see a sinusoidal signal. On short time scales the
amplitude of the sinusoidal appears constant. Over the length of a single five
minute measurement the amplitude varies with changes in the atmosphere. We only
need to extract the phase from the sine wave in order to determine the
polarization angles of the detectors. Before performing a fit to extract the
phase we normalize and then band-pass filter the raw detector time streams with a
0.5 Hz wide Butterworth filter centered on twice the grid rotation frequency.
The filtering damps low frequency atmospheric oscillations as well as high
frequency harmonics.

\begin{figure}[ht]
\begin{center}
\begin{tabular}{cc}
\includegraphics[width=0.5\linewidth]{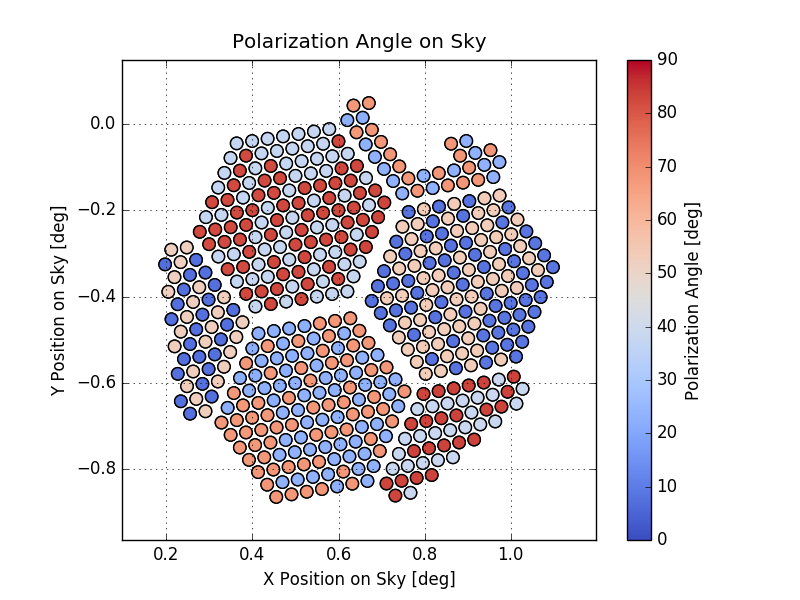} & \includegraphics[width=0.5\linewidth]{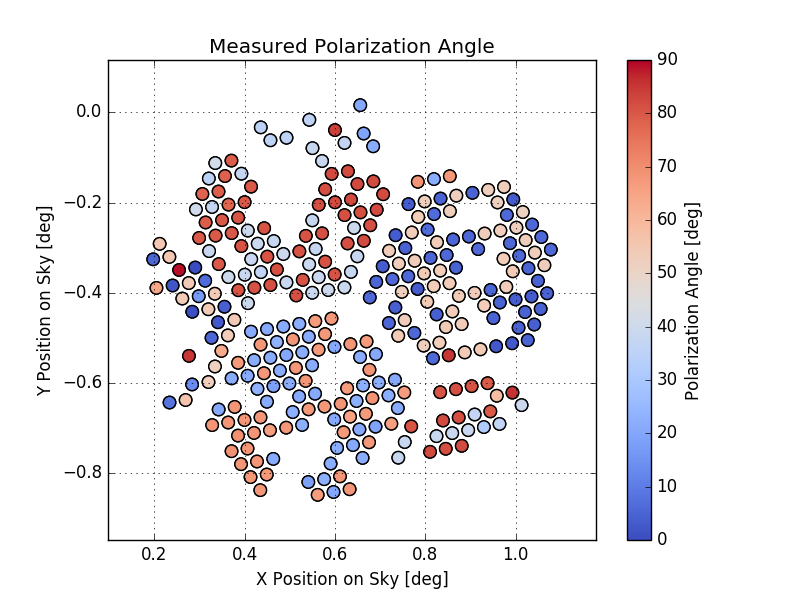}\\
\end{tabular}
\end{center}
\caption{\label{fig:physical_and_measured_angles} 
(Left): Physical detector angles for each detector pair projected on the sky.
The color bar describes the angles of one of the detectors in each feedhorn, the
other detector being orthogonal to the one plotted. (Right): Measured
polarization angles from the polarization grid data sets. Like in the
left panel, each circle represents a feedhorn which contains a pair of
detectors orthogonal to one another. The color bar describes the angle of a single
detector in the feed. Feeds are missing here, causing gaps when
compared to the left panel, if they are missing a single detector in the pair
due to less than 100\% yield in the array.}
\end{figure}

We use the phase determined from the time stream fits to calculate the angle of
each detector. Two detectors that are $180^{\circ}$ out of phase are
orthogonal so the detector angles are calculated as their phase angle
divided by two. The coordinate system for the measured detector angles is set
by the angular position of the polarization grid at the
start of a time stream. Without knowledge of the angular position of the grid,
we globally rotate the determined angles relative to a single
reference detector's physical coordinate system determined by the optical
modeling calibration without additional rotations due to the optics from CODE
V. This zeros the measured angle for that single detector and makes all other
measured angles relative to the chosen detector's local coordinate system.
Figure \ref{fig:physical_and_measured_angles} shows the physical angles
of the detectors on the sky and the measured angles from the time stream
fits.

\begin{figure} [ht]
\begin{center}
\begin{tabular}{cc}
\includegraphics[width=0.5\linewidth]{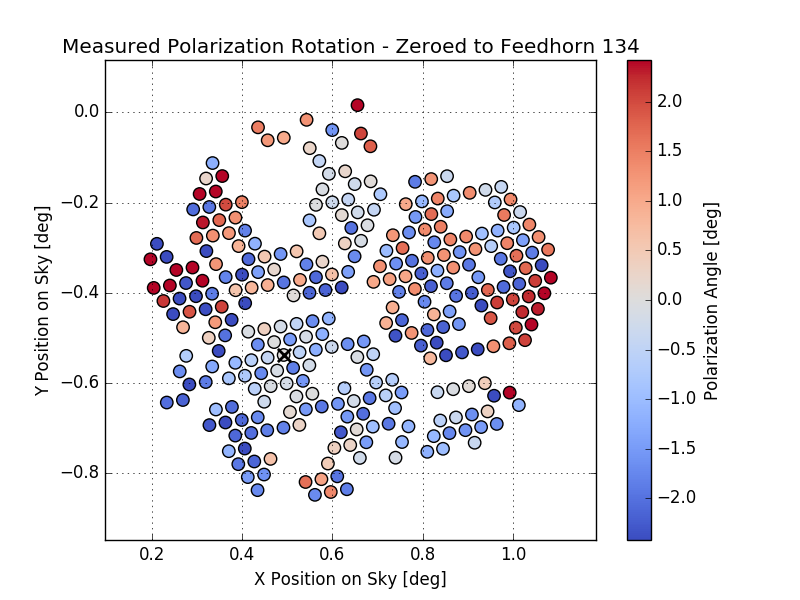} & \includegraphics[width=0.5\linewidth]{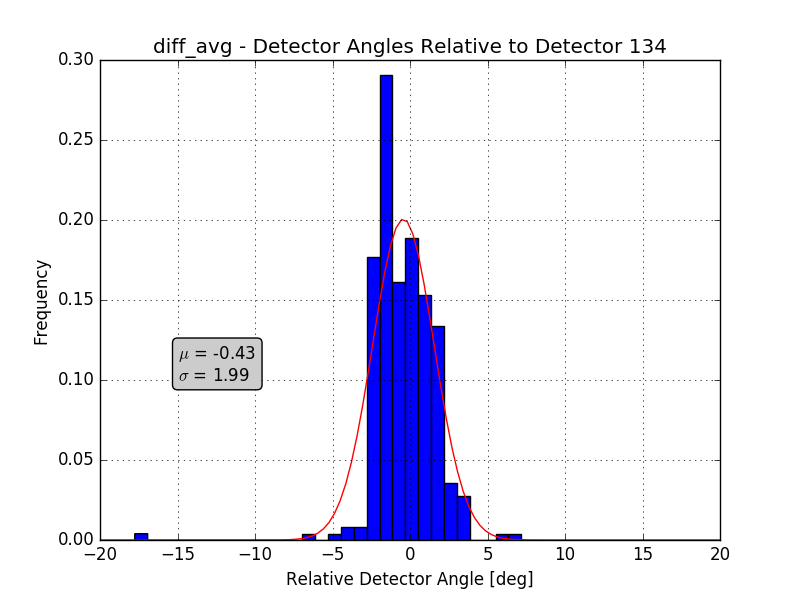}\\
\end{tabular}
\end{center}
\caption{\label{fig:polgrid_cal} 
(Left): Measured polarization rotations relative to detector 134, marked with
an 'X' on this plot. (Right): Histogram of average detector angles relative to
detector 134 and its colocated partner, which are used as a reference.}
\end{figure}

We then subtract the measured relative angles with the
physical angles of the detectors in the receiver. This produces a measurement
of polarization rotation due to the optics. This can be illustrated by
subtracting the two plots in Figure
\ref{fig:physical_and_measured_angles}, producing a plot similar to the left
plot of Figure \ref{fig:polgrid_cal}. Based on modeling in CODE V the
polarization rotation appears to be independent of input polarization angle, so
we then average the measured polarization rotation determined for colocated
detectors, which gives a measured polarization rotation per
feedhorn. We again rezero by removing the averaged phase for our selected
feedhorn from all feeds. The polarization angles can then be explored by
selecting different zeroing detectors across the focal plane.

The results for a single chosen detector are shown in Figure
\ref{fig:polgrid_cal}. For the selected reference detector coordinate system the
average difference from the physical angles for all the detectors across the
focal plane is $-0.4 \pm 2$ degrees. The mean is dependent
on which detector we choose as a reference, while the standard deviation,
reported as the error, is independent of the choice of reference.

There are still improvements to be made with the polarization grid
measurements. When fabricated, the grid was ablated in
small, roughly 1'' by 1'' square sections. While many sections are aligned,
some are misaligned by as much as $\sim25\,\mu\mathrm{m}$. This means the wires are
not uniformly straight across all areas of the grid. It is evident that the
uniformity of the grid greatly affects our ability to determine the detector
angles. Defects in the wire alignment that are radially
separated by $90^{\circ}$ may lead to the observed differences in the measured
relative angles of colocated detectors, which show minimal differences in
modeling. 

Defects in the grid can also arise in mounting.
Imperfections in mounting manifest themselves as wrinkles across the grid.
Coupled with small vibrations from the surroundings (i.e. wind, the HWP motor)
these wrinkles affect the signal propagating to the detectors, limiting our
ability to determine the polarization angles. Efforts to improve the
uniformity of the grid in fabrication and mounting are ongoing.

\section{COMPARISON TO EB NULLING}
The polarization angle calibration presented in Section
\ref{sec:optical_modeling} is the approach used in Naess et al. 2014 and for
all ACTPol results thus far \cite{2014JCAP...10..007N}. After implementing this
calibration approach, Naess et al. 2014 tested the polarization angle
calibration by checking the EB nulling angle and found it to be $\delta
\gamma_p = -0.2^\circ \pm 0.5^\circ$ \cite{2014JCAP...10..007N}. Without the
polarization angle calibration corrections presented in Section
\ref{sec:optical_modeling}, degree scale corrections would have been required
to null EB in Naess et al. 2014. This suggests that the optical modeling
correctly captures the largest scale polarization angle corrections with a
measurement uncertainty of $\pm0.5^\circ$. Our best polarization grid
measurements thus far have achieved a scatter of $\pm2^\circ$.  We plan to
pursue better astrophysical measurements and polarization grid measurements to
understand the calibration and test the modeling more precisely.

\section{SUMMARY}
We have outlined the steps taken to calibrate the ACTPol detector polarization
angles. Working from the fabricated design angles we use planet
observations to determine the installation angle and match these to our
model for the optical distortions. Finally, we add the additional
polarization rotation determined from optical modeling. This calibration
technique has produced detector angles consistent with a global offset angle
from zero when compared to the EB-nulling offset angle and is currently
used to calibrate the ACTPol detectors.

A technique for measuring the polarization rotation due to the reimaging optics
is also described. We observe a rotating polarization grid, fit the detector
time streams to determine the detector angles and take the difference of this
measured angle with the expected detector angle. The measured angles
demonstrate the potential for this technique to constrain polarization rotation
present in the reimaging optics; however, improvements need to be made with
respect to the uniformity of the polarization grid
to achieve sub-degree measurements of individual detector angles.

\vspace{-0.05in}

\acknowledgments
 
The work of BJK, KPC, KTC, EG, CM, BLS, JTW, and SMS was supported by NASA
Space Technology Research Fellowship awards. This work was supported by the
U.S. National Science Foundation through award 1440226. The development of
multichroic detectors and lenses was supported by NASA grants NNX13AE56G and
NNX14AB58G. MDN acknowledges support from NSF award AST-1454881.

\bibliography{report} 
\bibliographystyle{spiebib}

\end{document}